# Is 'superstatistics' really 'super'?

J. Dunning-Davies,
Department of Physics,
University of Hull,
Hull HU6 7RX,
England.

j.dunning-davies@hull.ac.uk

**Abstract.**

Attention is drawn to problems associated with superstatistics, including the apparent lack of knowledge of previous work in statistical physics displayed by workers in this supposedly new field.

The apparent lack of communication between workers in statistical mechanics and information theory is both surprising and disturbing. It has led over recent years to the pouring of old wine into new bottles with labelling that makes the uninitiated believe that ideas taken from information theory and presented in statistical mechanics are new. No better example is afforded than that of the so-called 'Tsallis entropy'[1], which is actually a result in information theory produced by the Hungarian school in the late 1960's and well documented in the book by Aczél and Daróczy [2]. As far as this example is concerned, the only difference, and that a very slight one, is in the choice of normalising factor. However, in some ways even more surprising, is the use of some known formulae of statistical mechanics in statistical mechanics itself with no reference to the original work. Even if used for supposedly different purposes, citing original references is the accepted norm.

Superstatistics [3] purports to provide a description of fluctuations in the inverse temperature in 'driven, non-equilibrium systems'. According to Beck and collaborators, a system at thermal equilibrium (i.e. distributed canonically) would be described by the usual Boltzmann factor, $e^{-\beta E}$, where $E$ is the energy of the system and $\beta$ the characteristic reservoir parameter, the inverse temperature of the heat bath with which the system is in thermal contact. However, in 'driven, non-equilibrium systems', one can rightly expect fluctuations in $\beta$, and consequently the system will be characterised by an 'effective Boltzmann factor' given by the Laplace transform

$$B(E) = \int_0^\infty e^{-\beta E} f(\beta) d\beta \quad (1)$$

of a normalised probability density, $f(\beta)$, which describes fluctuations in the inverse temperature. Further, in the discussions of superstatistics, a distinction is drawn between 'type-A' and 'type-B' superstatistics [4]. This distinction is, in fact, something of a red herring since it converts the inverse density of states (which is really what $B(E)$ is) into a normalised probability density function (pdf), $p(E)$, by incorporating the normalising factor $Z(\beta)$ into the definition of the pdf $f(\beta)$ so that it becomes $\tilde{f}(\beta) = f(\beta)/Z(\beta)$. The problem here is that neither Beck nor his collaborators was first to conceive of fluctuations in inverse temperature but they do not reference any earlier work on this. When it is realised that $\beta$ is an estimable parameter - that is, one to be estimated in terms of measurements made on the energy of the system - then it becomes apparent that any estimator, $\beta(E)$, of the inverse temperature, which must be a function of the energy, must fluctuate itself [5].

If $\bar{E}$ denotes the mean sample energy, Beck's formula (1) in the latest publication on superstatistics [6] is none other than equation (4.102) of Lavenda's 1991 book [5], where

$$B(\bar{E}) = e^{-S(\bar{E})},$$

and

$$f(\beta) = e^{-L(\beta)}.$$

Lavenda identified $L(\beta)$ as the logarithm of the moment generating function for the central moments of the energy. In a completely symmetrical picture, $S(E)$ is the logarithm of the moment generating function of moments in the inverse temperature. Lavenda terms this the 'dual' representation [5, p. 208] made possible by Bayes' theorem of inverse probability. The distinction made was that, whereas $E$ may be interpreted as a random variable in the limit-of-frequency sense, $\beta$ must be interpreted in the sense of degree-of-belief that certain values of $\beta$ are more likely than others. These ideas actually go back to Szilard [7] and Mandelbrot [8], as is documented in Lavenda's monograph [5].

Further, Lavenda emphasises the fact that, in the thermodynamic limit as Boltzmann's constant tends to zero, Laplace's method of evaluating (1) assumes the main contribution to come from a neighbourhood of $\beta_E$, in Beck's notation, which is the only minimum of
$$\beta E + L(\beta).$$
This effectively reduces the Laplace transform, (1), to the Legendre transform [5, eq.(4.84)]
$$L(\hat{\beta}) = S(\overline{E}) - \hat{\beta}(\overline{E}), \quad (2)$$
where $\hat{\beta}(\overline{E})$ is the best estimator available for the inverse temperature which is a function of the sample mean energy. Supposedly, this is contained in formulae (7)
$$\beta_E = \sup_{\beta}\{-\beta E + \ln f(\beta)\},$$
and (9)
$$\sup_{\beta}\{-\beta E + f(\beta)\} = -\beta_E E + \ln f(\beta_E),$$
of Beck's most recent contribution to superstatistics [6]. On comparing the two formulae, it is obvious something is indeed amiss.

Finally, it should be noted that even Beck's primary example is not new. His so-called $\chi^2$ pdf of $n$ degrees of freedom in his formula (12),
$$f(\beta) = \frac{1}{\Gamma(n/2)}\left(\frac{n}{2\beta_0}\right)^{n/2} \beta^{n/2-1} e^{-n\beta/2\beta_0},$$
which he attributes to Wilk and Wlodarczyk [9] and himself [10], may, once again, be found in Lavenda's 1991 monograph [5, eq.(4.97)]. All that is necessary to make the comparison is the simple change of notation $n = 3N$ and $n/2\beta_0 = \overline{E}$.

Other criticisms of actual superstatistics may be found in various references [11] but the main point being raised here is the need for people to be aware of, and reference, results already in existence in their field.